\documentclass[prd,preprintnumbers,aps,preprint,amsmath,amssymb,superscriptaddress,floatfix,nofootinbib,unsortedaddress]{revtex4}
\pdfoutput=1

\usepackage{amssymb, bm, amsmath, graphicx, physymb}
\usepackage[colorlinks=true,linktocpage=true,linkcolor=blue,citecolor=blue]{hyperref}
\usepackage[usenames,dvipsnames]{color}
\usepackage{slashed}
\usepackage[utf8]{inputenc}
\usepackage[T1]{fontenc}
\usepackage{enumerate}
\usepackage{amsfonts}
\usepackage{mathtools}
\usepackage{latexsym}
\usepackage{hyperref}
\usepackage{epstopdf} 
\usepackage{bbm}
\usepackage{soul}
\usepackage[normalem]{ulem}
\usepackage{braket}
\usepackage{nccmath}
\usepackage{graphicx}
\usepackage{subfigure}

\usepackage[e]{esvect}

\def\cN{{\cal N}}
\def\cR{{\cal R}}
\def\cI{{\cal I}}
\def\cO{{\cal O}}

\def\cQ{{\cal Q}}

\def\beps{{\boldsymbol \epsilon}}

\newcommand{\secn}[1]{Section~1}
\newcommand{\appn}[1]{Appendix~1}

\long\def\comment#1{ }

\def\and{\quad\text{and}\quad}

\def\q{{\boldsymbol q}}
\def\0{{\boldsymbol 0}}
\def\1{{\boldsymbol 1}}
\def\p{{\boldsymbol p}}
\def\l{{\boldsymbol l}}
\def\k{{\boldsymbol k}}
\def\x{{\boldsymbol x}}

\def\u{{\boldsymbol u}}

\def\b{{\boldsymbol b}}
\def\0{{\boldsymbol 0}}

\newcommand{\tx}{\text{x}}

\renewcommand\a{\alpha}
\renewcommand\b{\beta}
\renewcommand\d{\delta}
\renewcommand\l{\lambda}

\renewcommand\u{\upsilon}

\renewcommand\o{\omega}
\newcommand\e{\epsilon}
\newcommand\m{\mu}
\newcommand\n{\nu}

\newcommand\s{\sigma}

\def\u{{\boldsymbol u}}

\renewcommand{\part}{{\rm part}}

\newcommand{\be}{\begin{equation}}
\newcommand{\ee}{\end{equation}}
\newcommand{\bes}{\begin{subequations}}
\newcommand{\ees}{\end{subequations}}
\newcommand{\bea}{\begin{eqnarray}}
\newcommand{\eea}{\end{eqnarray}}

\begin{document}

\title{Gluon radiation inside a flowing medium}

\author{Matvey V. Kuzmin}
\email[Email: ]{kuzmin.mv19@physics.msu.ru}

\affiliation {Faculty of Physics, Moscow State University, Moscow 119991, Russia}
\author{Xo{\'{a}}n Mayo L\'{o}pez}
\email[Email: ]{xoan.mayo.lopez@usc.es}
\affiliation{Instituto Galego de F{\'{i}}sica de Altas Enerx{\'{i}}as,  Universidade de Santiago de Compostela, Santiago de Compostela 15782, Galicia, Spain}

\begin{abstract}
We compute the spectrum of gluons emitted by a highly energetic quark inside a flowing QCD medium, focusing on the soft gluon limit at first order in the opacity expansion. Specifically, we derive the leading energy-suppressed corrections to the double differential final parton distribution, and show that they are substantial for both static and flowing matter. In particular, we demonstrate that the corrections due to the transverse flow become large even at moderate energies and flow velocities, affecting drastically both the shape and magnitude of the spectrum. We observe that the final transverse momentum of the emitted gluon tends to align along the flow direction, resulting in a non-trivial  azimuthal distribution. These results can be directly implemented into the estimation of multiple observables to get a better understanding of the jet-medium interaction processes in HICs.
\end{abstract}

\maketitle

\section{Introduction}

Over the past few decades, our understanding of Quantum Chromodynamics (QCD) at extreme energies and densities has drastically improved, specially due to the high-energy heavy-ion collision (HIC) experiments taking place at RHIC and LHC, for a review see e.g. \cite{Busza:2018rrf,Cunqueiro:2021wls, Apolinario:2022vzg}. In HIC experiments, a new extreme state of nuclear matter is created far from the equilibrium, undergoing a multiphase evolution. After the thermalization into a nearly ideal liquid, known as the quark-gluon plasma (QGP), the matter continues expanding and cooling down until it reaches the hadronization transition and it is observed by the detectors as a gas of hadrons. The observation of the formation of the QGP stands as a major breakthrough in our understanding of the QCD phase diagram. More recently, the community has focused on characterizing the dynamical properties  of the matter created in HIC experiments. One of the main instruments to achieve this goal are jets: collimated sprays of particles formed through the branching of a highly energetic parton originated at the initial hard scattering in the nucleus collision. Jets travel through the medium interacting with it throughout the different stages of its evolution. Thus, medium-induced modifications of jet properties encode information about the dynamics of the nuclear matter and can be used to probe of the QGP, see e.g. \cite{Vitev:2002pf, Wang:2002ri, JET:2013cls, Betz:2014cza, Xu:2014ica, Djordjevic:2016vfo, Apolinario:2017sob, Du:2021pqa,Sadofyev:2021ohn}. 

The interaction of energetic partons with the nuclear matter can be described within perturbative QCD (pQCD).  Under this approach, the medium is commonly modelled by a background stochastic color field, averaging  all the observables over its possible configurations, see e.g. \cite{Gyulassy:1993hr, Zakharov:1996fv, Baier:1996kr,  Wiedemann:2000ez, Wiedemann:2000za, Gyulassy:2000er, Gyulassy:2002yv, Arnold:2002ja}. In this formalism, partons get deflected interacting with the medium field and lose energy, mostly through soft gluon emissions. Multiple simplifications are needed to make the calculations tractable in this framework. For instance, the medium is usually assumed to be static and transversely homogeneous with a finite longitudinal extension\footnote{Longitudinal and transverse directions are defined with respect to the initial momentum of the leading parton}. Additionally, the so-called eikonal approximation, which assumes the energy of the initial parton $E$ to be much larger than the characteristic transverse momentum scale $\bot$, is frequently used. As a result of these simplifications, the theoretical description of the jet decouples from the medium evolution, and the imaging of the different stages of the QGP becomes impossible, see e.g. the discussion in \cite{Sadofyev:2021ohn}.

There were early attempts to study the modifications of the jet properties due to the effect of the medium flow, see e.g. \cite{Gyulassy:2000gk, Gyulassy:2001kr,Baier:1998yf,Baier:2006pt, Liu:2006he, Renk:2006sx,Armesto:2004pt, Armesto:2004vz}. However, most of these works are based on phenomenologically-motivated models, rely on strong simplifying assumptions or focus solely on longitudinal expansion. Only more recently, multiple formal extensions of the jet-medium interaction framework have been presented, accounting for inhomogeneity and flow of the nuclear matter in pQCD calculations\footnote{Notice that there were developments on the theoretical description of probe-medium interactions for strongly coupled evolving plasmas in holography, see e.g. \cite{Lekaveckas:2013lha, Rajagopal:2015roa, Sadofyev:2015hxa, Reiten:2019fta, Arefeva:2020jvo,Liu:2006he,Nijs:2023dbc}.}, see e.g. \cite{He:2020iow,Sadofyev:2021ohn,Ipp:2020mjc, Hauksson:2021okc,Carrington:2021dvw,Antiporda:2021hpk,Sadofyev:2022hhw,Barata:2022krd,Fu:2022idl,Barata:2022utc,Hauksson:2023tze,Boguslavski:2023alu,Barata:2023qds,Barata:2023zqg,Andres:2022ndd,Barata:2024xwy} and references therin. In this paper, we continue developing the theoretical formalism by deriving the medium-induced soft gluon spectrum in the presence of medium flow at first order in the opacity expansion. The calculation is done for spin-1 gluons of the underlying gauge theory while keeping the first subeikonal corrections, which are non-zero even in the case of static matter, going beyond the discussion in \cite{Sadofyev:2021ohn,Kuzmin:2023hko}. We show that the flow modifications to the spectrum can be substantial and required for an accurate characterization of the QGP evolution.

\section{Theoretical formalism and setup}

In this work, we study the medium-induced gluon spectrum sourced by a highly-energetic quark interacting with flowing nuclear matter created in HIC within the opacity expansion\footnote{The opacity expansion is a perturbative expansion which can be related with the expected number of scatterings that a high energy particle will undergo in the medium before escaping. The formalism used here was introduced by Gyulassy, Levai and Vitev, see e.g. \cite{Gyulassy:2000fs,Gyulassy:2000er}. There are other theoretical frameworks based on the same perturbative approach, such as the resummed formalism  introduced by Baier, Dokshitzer, Mueller, Peigne, Schiff, and Zakharov, see e.g. \cite{Zakharov:1996fv, Baier:1996kr}}. We extend the formalism developed in \cite{Sadofyev:2021ohn} to the case of a spin-1 gluon of the underlying gauge theory, ignoring the energy-suppressed effects of the quark spin, which appear at higher subeikonal orders, thus working with scalar quarks in the fundamental representation. 

The effect of the thermal matter on jet particles is modelled by a classical stochastic color field generated by moving quasi-particle sources. We assume the in-medium sources to be massive, essentially treating them as classical currents, and ignore the medium response and collisional energy loss. Thus, the background field can be written as 
\begin{equation}
    \label{Amu}
    gA^{a\mu}_{\text{ext}}(q) =  \sum_i \, u^\mu \, e^{-i(\q\cdot\x_i + q_z z_i)} \, t^a_i \, v(q) \; (2\pi)\, \delta(q_0-\q\cdot\u_i-q_z u_{zi})\, ,
\end{equation}
where $u_\mu=(1, \u, u_z)$ is the four-velocity of the sources with the relativistic $\gamma$-factor removed, $(\x_i,z_i)$ is the position of the $i$-th source, $t^a_i$ its color charge projection onto the $SU(N_c)$ generators, and the sum runs over all the sources in the medium. The single-source scattering potential $v(q)$ controls the interaction of the highly energetic parton with the in-medium quasi-particles. For simplicity, we will assume that the nuclear matter is homogeneous on the transverse directions to the jet, therefore neglecting the possible gradient corrections introduced by the transverse structure of the matter, and focusing solely on the correction due to the presence of local flow. For a detailed discussion of the gradient effects,  see e.g. \cite{Barata:2022krd,Barata:2022utc,Barata:2023qds,Kuzmin:2023hko,He:2020iow,Hauksson:2021okc}.

The form of the field in \eqref{Amu} is obtained solving the classical field equation in the Lorenz gauge $\partial_\m A^\m = 0$. However, it is convenient to work in the axial gauge $n\cdot A=0$, in which only the two physical gluon polarizations appear. The two different gauge conditions can be satisfied by \eqref{Amu} at the same time by considering a reference frame\footnote{One should notice that boosting on any transverse direction mixes the energy of the particle and the transverse components of its momenta, breaking the eikonal expansion. However, longitudinal boosts are allowed and enable us to obtain the $u_z$ corrections directly from our final result.} in which $u_z = 0$, making $A_z = 0$. This way, choosing $n_\mu = (0,\boldsymbol{0},1)$ ensures that the field satisfies $n\cdot A = A_z = 0$.

The specific form of the single-source scattering potential is model dependent, and there are several options discussed in the literature, see e.g. the discussion in  \cite{Antiporda:2021hpk,Barata:2020rdn,Caron-Huot:2010qjx}. Even though most of the discussion in this paper is general for any screened scattering potential,  we will consider the Gyulassy-Wang (GW) model \cite{Gyulassy:1993hr} to make the results explicit. The potential reads
\begin{equation}
    v(q)=\frac{g^2}{q^2 -\mu^2}\, ,
\end{equation}
where $g$ is the effective in-medium strong coupling, and $\mu$ is a screening scale (Debye mass) controlled by the temperature of the matter. We assume the medium to be diluted and extended enough so that $\m(z_i-z_0)\gg 1$. Therefore, the contribution of the poles coming from the scattering potential are exponentially suppressed and can be neglected.

As it is usually done in pQCD jet calculations, the stochastic color field is assumed to have Gaussian statistics, and therefore only pairwise averages are non-zero. Furthermore, we assume that the jet-medium interactions are sufficiently local and there are no correlations between different sources. Thus, the averaging over the field configurations enforces 
\begin{equation}
    \langle t^a_i t^b_j \rangle = \frac{1}{2N_c} \d^{ab} \d_{ij} \, ,
\end{equation}
where in-medium sources have been assumed to be in the fundamental representation of $SU(N_c)$.

The energy of the initial parton $E$ is assumed to be much larger than the characteristic transverse momentum scale $\perp$ and the Debye mass, following the eikonal expansion on powers of $E$. It is convenient to introduce the energy fraction carried by the gluon $\text{x}=\frac{\o}{E}$, where $\o$ is the total energy of the emitted gluon. We will consider the soft gluon limit, i.e. we take $\text{x}\ll 1$, keeping only the subeikonal terms enhanced by inverse powers of x up to $\mathcal{O}\left(\frac{1}{\text{x}E}\left(\frac{\perp^2}{\text{x}E} z\right)^{\text{n}}\right) $ with $\text{n}\ge0$, and neglecting any possible correction beyond that. One should notice that, since the Landau-Pomeranchuk-Migdal (LPM) phases are also enhanced by the length of the medium, they involve terms scaling as $\left(\frac{\perp^2}{\text{x}E} z\right)^{\text{n}} \sim \cO\left(1\right)$ and must be kept to all orders.

In the axial gauge that we are using, the gluon propagator is transverse to the axial vector $n$ and the gluon 4-momentum $k$. It can be written as  
\begin{equation}
    G_{\m\n}(k)=\frac{-i N_{\m\n}(k)}{k^2 + i\e}\, ,
\end{equation}
where the numerator is expressed in terms of these two vectors:
\begin{align}
     N_{\mu\nu}(k)=g_{\mu\nu}+n^2 \frac{k_\mu k_\nu}{(k\cdot n)^2-k^2 n^2} + k^2 \frac{n_\mu n_\nu}{(k\cdot n)^2-k^2 n^2}
    -(k\cdot n)\frac{k_\mu n_\nu + k_\nu n_\mu}{(k\cdot n)^2 -k^2 n^2} \, .
\end{align}
The normalized polarization vector, which satisfies $n\cdot \e^*(k)=0$ and $k\cdot\e^*(k)=0$ in this gauge, is given by 
\begin{equation}
    \e_\mu(k) = \left(\frac{\beps\cdot\k}{\o},\beps,0\right) \, ,
\end{equation}
with $\beps$ being the transverse polarization vector, $\k$ being the final transverse momentum of the gluon, and the polarization index being omitted. Even though the transverse polarization is in general a function of the whole 4-momentum $\beps\equiv\beps(k)$, we see that $\sum_\l \beps^\l_\a \beps^\l_\b = \d_{\a\b} + \mathcal{O}\left(\frac{1}{(\tx E)^2}\right)$, with $\a$ and $\b$ running in the 2d transverse plane and the correction to the delta being beyond the accuracy of our discussion.

Finally, we will take the so-called broad source approximation, where the initial source  $J$, which carries the information about production of the initial energetic parton, is assumed to have a sufficiently weak dependence on the transverse components of the momentum.

\section{Medium-induced gluon spectrum}

\begin{figure}[t!] 
    \centering
    \includegraphics[width=0.7\textwidth]{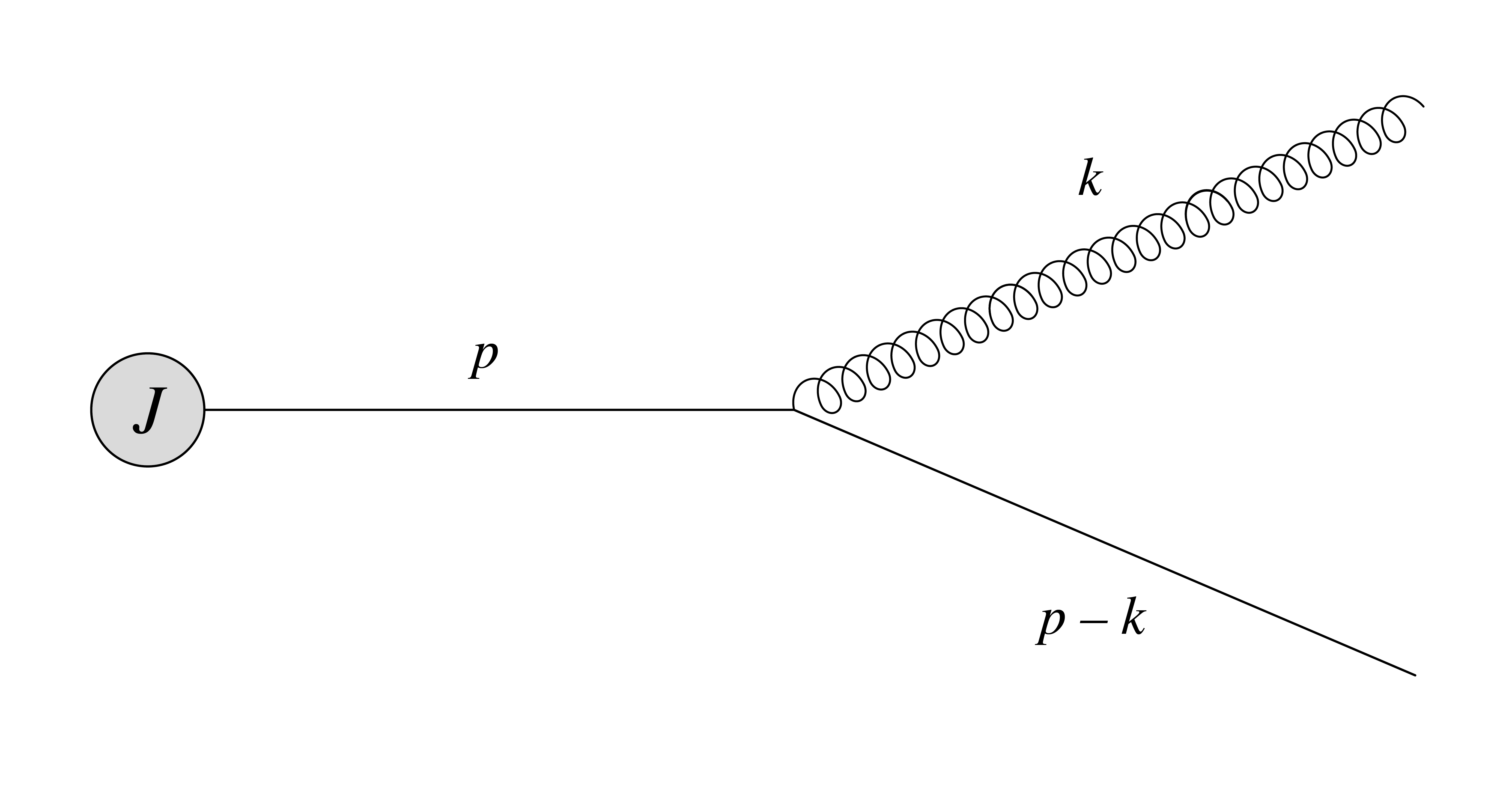}
    \vspace*{-10mm}\caption{The vacuum vertex of gluon emission from a quark. It corresponds to the zeroth order in the opacity expansion.}
\label{f:R0}
\end{figure}
\begin{figure}[t!] 
    \centering
    \includegraphics[width=1\textwidth]{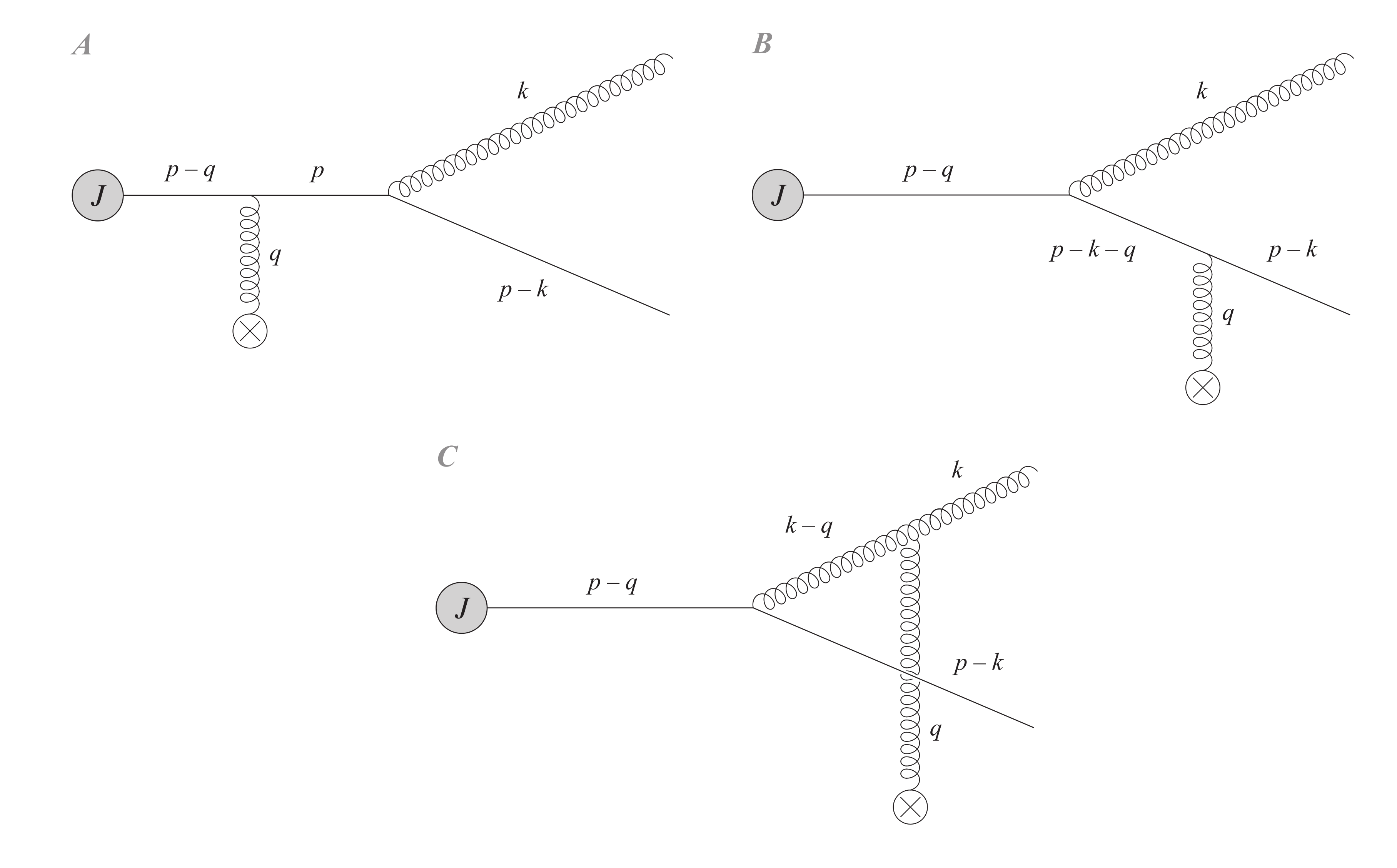}
    \vspace*{-10mm}\caption{The three single-Born diagrams contributing to the medium-induced gluon emission at the first order in the opacity expansion. }
\label{f:SB diagrams}
\end{figure}
\begin{figure}[t!] 
    \centering
    \includegraphics[width=1\textwidth]{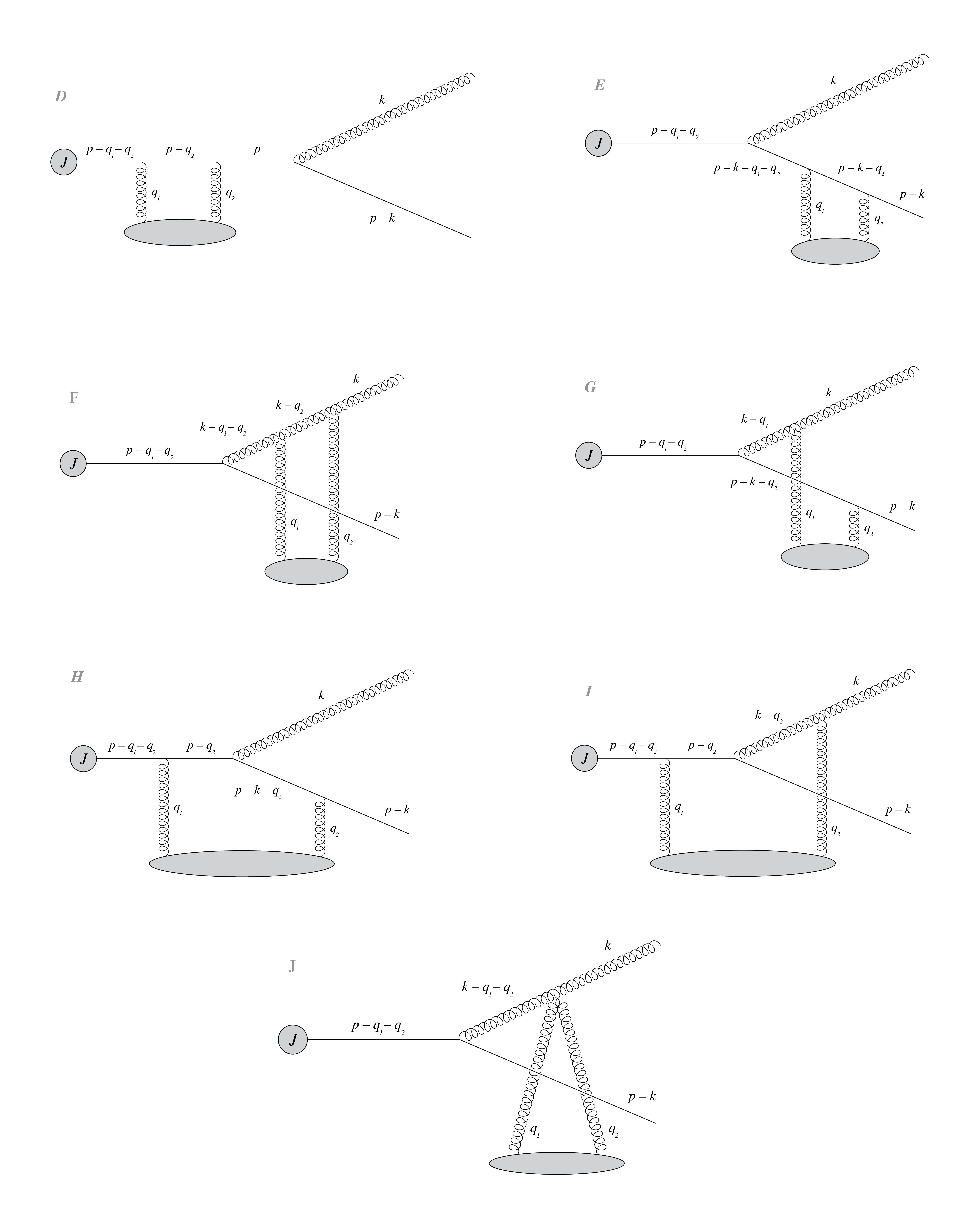}
    \vspace*{-10mm}\caption{The seven double-Born diagrams contributing to the medium-induced gluon emission at the first order in the opacity expansion. }
\label{f:DB diagrams}
\end{figure}

In this section, we compute the medium-induced spectrum for a highly energetic parton traversing a flowing medium and radiating a soft gluon up to the first order in the opacity expansion. Let us start by considering the branching of the initial quark into an on-shell quark with momentum $(p-k)_\m$ and an on-shell gluon with momentum $k_\mu$ in the vacuum. This diagram, shown in Fig.~\ref{f:R0} at the zeroth order in the opacity expansion,  reads
\begin{align}
    i\cR_0=\left[i\, g\, t^r_{proj} \, (2p-k)_\mu \,\epsilon^{*\mu}(k)\right] \, \frac{i}{p^2+i\epsilon}  \, J(p) \simeq - g\, t^r_{proj} \, \frac{2 \, \beps \cdot\k}{\k^2} \, J(p) \, ,
\end{align}
where $t^r_{proj}$ is the color generator of the energetic quark (`projectile') from the emission vertex, with $r$ being the color of the final gluon, and $J(p)$ is the initial source of the leading parton. Medium-induced radiation is sometimes discussed in the literature in terms of ligth-front wave function of the elementary branchings. In particular, the light-front wave function of the splitting in the vacuum can be read from the latter equation, being $\psi(\tx,\k) = -\frac{2\, \beps\cdot\k}{\k^2}$ .

At first order in the opacity expansion $N=1$, there are 10 different diagrams, shown in Fig.~\ref{f:SB diagrams} and Fig.~\ref{f:DB diagrams}, that contribute to the amplitude of this process: three single-Born (SB) diagrams and seven double-Born (DB) diagrams. In what follows, we will study them all following closely the procedure in \cite{Sadofyev:2021ohn,Kuzmin:2023hko}.

\subsection{Single Born contributions}

\subsubsection{The amplitude level}

Let us start with the simplest contribution to the amplitude, $\mathcal{R}_A$, where the branching occurs after the scattering of the background field. This amplitude reads
\begin{align}
\label{RA beginning}
    i\cR_A & =\int\frac{d^4q}{(2\pi)^4}\, [i\,g\, t^r_{proj} \,(2p-k)_\nu \,\epsilon^{*\nu}(k)] \, \frac{i}{p^2+i\epsilon} \, 
    [i \, t^a_{proj} \, (2p-q)_\mu \,g A^{a\mu}(q)] \, \frac{i}{(p-q)^2+i\epsilon} \, J(p-q) \notag
    \\ & \hspace{0.5cm} =\frac{2\,\beps \cdot \k }{\k^2} \, \sum_i \, t^r_{proj}  t^a_{proj}t^a_i \, \int_q \, e^{-i(\q_i\cdot\x_i + q_z z_i)} \frac{(2\,g\,E)\,v(q)}{(p-q)^2+i\epsilon} \, J(p-q) \, , 
\end{align}
where we have performed $q_0$ integration using the delta function in \eqref{Amu}, and $J(p-q)$ is assumed to be produced by a hard scattering event at $\x_0=0$ and $z_0=0$. We have also used a shorthand notation for integrals running over the whole three-dimensional space $\int_x\equiv\int d^3 x$ (and $\int_k\equiv\int \frac{d^3k}{(2\pi)^3}$), and over the two-dimensional transverse space $\int_{\x}\equiv \int d^2\x$ (and $\int_\k\equiv\int \frac{d^2\k}{(2\pi)^2}$).

In order to perform the integration over $q_z$, we assume that $J$ is slowly varying, so that only four poles, two coming from the propagator and two coming from the scattering potential, can contribute to the integral. As we have already discussed in the previous section, we assume the medium to be diluted and extended enough, with the poles coming from $v(q)$ being exponentially suppressed. Consequently, the integral is controlled by the two poles of the quark propagator, which are
\begin{subequations} \label{Qpq poles}
\begin{align}
     \cQ_{p-q}^+ &\simeq 2E \label{QMpq}\,  , 
     \\  \cQ_{p-q}^- &\simeq \q\cdot\u - \frac{\k^2}{2 \tx E} \label{Qmpq} \, ,
\end{align}
\end{subequations}
keeping corrections to the leading order (LO) terms up to $\cO\left(\frac{1}{\tx E}\right)$.
Thus, we can rewrite the initial-state scattering amplitude as
\begin{align}
    i\cR_A & = -ig \, \frac{2\beps\cdot\k}{\k^2} \, \sum_i t^r_g t^a_{proj} t^a_i \int_\q e^{-i\q\cdot\x_i} e^{-i\cQ^-_{\p-\q}z_i} \theta(z_i) \, \notag
    \\  & \times  \left(1-\frac{\k^2 \, \u\cdot\q}{\text{x}E} \frac{1}{v(\q)} \frac{\partial v(\q)}{\partial \q^2}\right) v(\q) J(E,\p-\q) \, ,
\end{align}
where we have expanded the scattering potential and the initial source to our working accuracy after the integration, and $v(\q)\equiv-\frac{g^2}{\q^2+\mu^2}$ has been introduced.

The next diagram corresponds to the final-state quark interacting with the medium, and its contribution to the amplitude reads
\begin{align}
    i\cR_B &= \int \frac{d^4q}{(2\pi)^4}\, [i\, t^a_{proj}\, (2(p-k)-q)_\mu \, gA_{ext}^{a\mu}] \, \frac{i}{(p-k-q)^2+i\epsilon} \,\notag
    \\ & \hspace{2cm} \times [i \,g \,t^r_{proj}  (2(p-q)-k)_\nu \, \epsilon^{*\nu}(k)] \, \frac{i}{(p-q)^2+i\epsilon} \, J(p-q) \, .
\end{align}
Using the constraint from the field to integrate over $q_0$  and performing $q_z$ integration as it has been done for $\cR_A$, this amplitude can be written as 
\begin{align} \label{Rb amplitude}
    i\cR_B &= i g \, \frac{2\beps\cdot\k}{\k^2} \, \sum_i t^a_{proj} t^a_i t^r_g \int_\q e^{-i\q\cdot\x_i} \theta(z_i)  \, \notag
    \\ & \times\bigg[e^{-iQ^-_{p-q}z_i} \left(1-\frac{\k^2 \, \u\cdot\q}{\tx E} \frac{1}{v(\q)} \frac{\partial v(\q)}{\partial \q^2}\right) - e^{-iQ^-_{p-k-q}z_i}\bigg] v(\q) J(E,\p-\q) \, ,
\end{align}
where the poles of the second propagator relevant for this diagram are 
\begin{subequations}
\begin{align}
    &\cQ^+_{p-k-q} \simeq 2E \, 
    \\ &\cQ^-_{p-k-q} \simeq \q\cdot\u \, .
\end{align}
\end{subequations}

The last SB contribution to the amplitude of the medium-induced spectrumm at $N=1$ corresponds to the final-state gluon interacting with the nuclear matter. It reads
\begin{align}
    i\cR_C &= \int \frac{d^4 q}{(2\pi)^4} \, i(2p-k-q)_\mu \, t^a_{proj} \, \frac{-i N^{\mu\nu}(k-q)}{(k-q)^2+i\epsilon} \,\Gamma^{abc}_{\nu\alpha\rho}(k-q,q,-k) \notag\,
    \\ &\hspace{3.5cm} \times  
    gA^{b \alpha}(q) \, \epsilon^{*\rho}(k) \, \frac{i}{(p-q)^2+i\epsilon}  \, J(p-q) \, ,
\end{align}
with $\Gamma^{abc}_{\nu\alpha\rho}(k-q,q,-k)$ being the three-gluon vertex, defined as 
\begin{align}\label{3gluonvertex}
    \Gamma^{abc}_{\mu\nu\rho}(k,p,q) = g\,f^{abc}\,[g_{\mu\nu} (k-p)_{\rho} + g_{\nu\rho} (p-q)_\mu + g_{\rho\mu} (q-k)_\nu] ,
\end{align}
where all the momenta go towards the vertex. As it was discussed in the previous section, $N^{\mu\nu}$, $\epsilon^*_\mu$, and $A^{a\mu}_{\text{ext}}$ are all transverse to $n_\mu=(0, \,\boldsymbol{0}, \,1)$, so the product of these three objects and \eqref{3gluonvertex} is independent of the $z$-component of the momentum, and no new poles can arise from it. Hence, taking both $q_0$ and $q_z$ integrals as in the previous cases, $\cR_c$ can be written as 
\begin{align}
    i \cR_C &= g \, \sum_i  f^{abc} t^a_g t^b_i \int_\q \theta(z_i) e^{-i\q\cdot\x_i} \left(2\frac{\beps\cdot(\k-\q)}{(\k-\q)^2} - 2 \,\frac{\beps\cdot\k\, (\k-\q)\cdot\u-\beps\cdot\u \, (\k-\q)\cdot\q}{(\k-\q)^2 \, \tx E} \right) \, \notag
    \\ & \hspace{1cm} \times  \bigg[e^{-iQ^-_{p-q}z_i} \left(1-\frac{\k^2 \, \u\cdot\q}{\tx E} \frac{1}{v(\q)} \frac{\partial v(\q)}{\partial \q^2}\right)  \, \notag
    \\ & \hspace{2cm} - e^{-iQ^-_{k-q}z_i} \left(1-\frac{\u\cdot\q \, (\k^2-(\k-\q)^2)}{\tx E}\frac{1}{v(\q)} \frac{\partial v(\q)}{\partial \q^2}\right) \bigg]  v(\q) J(E,\p-\q) \, ,
\end{align}
where the poles of the gluon propagator entering in this diagram are
\begin{subequations}
\begin{align}
    & \cQ^+_{k-q}\simeq 2 \tx E \left(1-\frac{\q\cdot\u}{2xE}\right) \, , 
    \\ & \cQ^-_{k-q}\simeq \q\cdot\u + \frac{(\k-\q)^2-\k^2}{2 \tx E} \, .
\end{align}
\end{subequations}

\subsubsection{SB contributions to the squared amplitude}

In order to get the SB contribution to the squared amplitude of the process at $N=1$, we must average over initial quantum numbers, sum over final ones and perform the medium averages. In addition to squaring each diagram individually, we have to take into account the interference terms.

Let us start by squaring $\cR_A$, which can be written as 
\begin{align} \label{R^2_A initial}
    \langle|\cR_A|^2\rangle & = g^2 \, \frac{C_F^2}{2N_c}\, \frac{4}{\k^2}\sum_i \int_{\q,\bar{\q}} \theta(z_i) e^{-i(\q-\bar{\q})\cdot\x_i} e^{-i(Q^-_{p-q}-Q^-_{p-\bar{q}})z_i} \, \notag
    \\ & \times \left(1-\frac{\k^2\, \u\cdot\q}{xE}\frac{1}{v(\q)} \frac{\partial v(\q)}{\partial \q^2} -\frac{\k^2\, \u\cdot\bar{\q}}{xE}\frac{1}{v(\bar{\q})} \frac{\partial v(\bar{\q})}{\partial \bar{\q}^2}\right) v(\q)v(\bar{\q}) J(E,\p-\q) J^*(E,\p-\bar{\q}) \, ,
\end{align}
where the color matrices combine to an overall factor of $\frac{C_F^2}{2N_c}$.

For convenience, we will take the continuum limit for the sum over all scattering centers, replacing it by an integral over the number density of in-medium sources as 
\begin{align}
    \sum_i f_i = N\Big(\frac{1}{N}\sum_i f_i\Big) = N\langle f\rangle = \int_{\x,z} \rho(z) f(\x,z) \, .
\end{align}
Since the medium is assumed to be homogeneous in the transverse plane, nothing depends on the transverse position $\x$  apart from the Fourier factor. Thus, integrating over it one gets $(2\pi)\d^{2}(\q-\bar{\q})$, which allows for a trivial integration over $\bar{\q}$. After performing these two integrals, $\cR_A$ contribution to the squared amplitude at our working accuracy reads
\begin{align}
    \langle|R_A|^2\rangle &= \frac{C_F^2}{N_c}\, g^2 \, \frac{2}{\k^2} \int_{0}^{L}dz \,\rho(z) \int_{\q} \, \left(1-  \frac{\k^2\, \u\cdot\q}{\tx E}\frac{1}{v(\q)^2} \frac{\partial v(\q)^2}{\partial \q^2} \right) v(\q)^2 |J(E,\p-\q)|^2  \, ,
\end{align}
where we have assumed matter to have a finite longitudinal size $L$. Under the broad source approximation, we can average the integral in $\langle|R_A|^2\rangle$ over the angles, cancelling the second term inside the brackets. Thus, there are no subeikonal corrections coming from this contribution.

Following the same procedure done for $\langle|R_A|^2\rangle$, we can perform the averages, take the integrals over $\x$ and $\bar{\q}$,  and write $\cR_B$ squared as
\begin{align}
    \langle |\cR_B|^2\rangle &= \frac{C_F^2}{N_c} \, g^2 \, \frac{4}{\k^2}  \int_{0}^L dz \,\rho(z)  \int_{\q} \left(1-\cos\left( \frac{\k^2}{2\text{x}E} \, z\right)\right) v(\q)^2 |J(E,\p-\q)|^2\, , 
\end{align}
where the latter expression has already been averaged over the angles, cancelling the odd subeikonal flow corrections. The amplitude of diagram \eqref{Rb amplitude} has the contribution of two different poles, which results in the appearance of a cosine of the LPM phase after taking the square.

The last amplitude obtained in the previous subsection can be squared and averaged in the very same way, resulting in 
\begin{align} \label{R_c^2}
    & \langle |\cR_C|^2\rangle = C_F \,g^2 \, \int_0^L dz\, \rho(z) \int_{\q} \frac{4}{(\k-\q)^2} \left(1-2\frac{(\k-\q)\cdot\u}{\tx E}\right)\, \notag
    \\ & \hspace{0.5cm} \times \left(1 -\frac{\u\cdot\q \, (2\k^2 - (\k-\q)^2)}{2\text{x}E} \frac{1}{v^2}\frac{\partial v^2}{\partial \q^2}\right) \left(1 - \cos\left( \frac{(\k-\q)^2}{2\text{x}E} \, z\right)\right)  v(\q)^2  |J(E,\p-\q)|^2 \, .
\end{align}
As in the previous diagram, the contribution of two different poles results in the cosine of the LPM phase. The dependence of the phases on $\q$ in \eqref{R_c^2} results in two non-vanishing subeikonal flow corrections -- one comes as a global factor, while the other is a modification of the scattering potential. 

The three SB interference terms are computed following the very same procedure -- after averaging over initial quantum numbers one should integrate the expression over $\x$, assuming the thermodynamic parameters to be independent of the transverse coordinates, and getting $(2\pi)^2\d^{(2)}(\q-\bar{\q})$, which forces the transverse momenta exchanged with the medium in amplitude and conjugated amplitude to be the same. Doing so, we can write the first interference term as  
\begin{align}
    \langle \cR_A \cR^*_B \rangle + \text{c.c.} &=   \frac{C_F}{N^2_c} \, g^2 \, \frac{2}{\k^2} \int_0^Ldz\, \rho(z) \int_{\q} \left(1 - \cos\left(\frac{\k^2}{2\text{x}E} \, z\right) \right) v(\q)^2 |J(E,\p-\q)|^2 \, ,
\end{align}
which also coincides with the contribution in the absence of any medium flow. In turn, the other two interference terms do get modified due to the presence of the flow and have several non-vanishing subeikonal corrections. They read 
\begin{align}
    & \langle  \cR_A \cR_C^*\rangle + \text{c.c.} = -  C_F  \, g^2 \, \int_0^L dz \, \rho(z) \int_{\q}  \frac{2}{\k^2 (\k-\q)^2}  \, \notag
    \\ & \times \left( \k\cdot (\k-\q)-\k^2 \, \frac{(\k-\q)\cdot \u}{\text{x} E} +\k\cdot\u \, \frac{(\k-\q)\cdot\q}{\text{x}E} \right)  \left(1- \frac{\k^2\, \q\cdot\u}{\text{x}E} \frac{1}{v^2} \frac{\partial v^2}{\partial \q^2}\right) \, \notag
    \\ & \times \bigg[ 1 -  \left(1 + \frac{\q\cdot\u (\k-\q)^2}{2\text{x}E} \frac{1}{v^2} \frac{\partial v^2}{\partial \q^2}\right) \cos\left(\frac{(\k-\q)^2}{2\text{x}E}\, z\right)\bigg] \, v(\q)^2 |J(E,\p-\q)|^2  \, ,
\end{align}
and 
\begin{align}
    & \langle R_B R^*_C \rangle + \text{c.c.} = - C_F \, g^2 \, \int_0^L dz\, \rho(z) \int_{\q} \frac{2}{\k^2(\k-\q)^2}  \, \notag
    \\ & \times \left(\k\cdot(\k-\q)-\k^2\frac{(\k-\q)\cdot \u}{\text{x} E} +\k\cdot\u \frac{(\k-\q)\cdot\q}{\text{x}E} \right) \left(1- \frac{\k^2 \, \u\cdot\q}{\text{x}E} \frac{1}{v^2} \frac{\partial v^2}{\partial \q^2}\right) \, \notag
    \\ & \times \bigg[  1 - \left(1 + \frac{\q\cdot\u \,(\k-\q)^2}{2\text{x}E} \frac{1}{v^2} \frac{\partial v^2}{\partial \q^2}\right) \cos\left(\frac{(\k-\q)^2}{2\text{x}E}\, z\right) - \left(1 + \frac{\k^2 \, \q\cdot\u}{2\text{x}E} \frac{1}{v^2} \frac{\partial v^2}{\partial \q^2}\right) \cos\left(\frac{\k^2}{2\text{x}E}\, z\right)  \, \notag
    \\ & \hspace{1.0 cm}  + \left(1 + \frac{\q\cdot\u \, (\k^2 + (\k-\q)^2)}{2\text{x}E} \frac{1}{v^2} \frac{\partial v^2}{\partial \q^2}\right) \cos\left(\frac{(\k-\q)^2-\k^2}{2\text{x}E}\, z\right) \bigg] \, v(\q)^2 |J(E,\p-\q)|^2 \,.
\end{align}
It may be instructive to notice that the difference between the scalar and the spin-1 gluon emission for SB diagrams comes from the interaction of the gluon with the medium field in $\cR_c$. For real spin-1 gluons, the light-front wave function of this diagram is more complex and results in richer structures.

\subsection{Double Born diagrams}

Now, we turn to studying the seven DB contributions to the amplitude of the soft gluon emission, shown in Fig.~\ref{f:DB diagrams}. The diagram with a 4-gluon vertex is usually ignored in the literature since it does not contribute to the leading eikonal accuracy, see e.g. the discussion in  \cite{Sadofyev:2021ohn}. Nevertheless, the 4-gluon vertex diagram contributes at $\mathcal{O}\left(\frac{\perp^2}{\text{x}E} \left(\frac{\perp^2}{\text{x}E}z\right)^n\right)$, and must be taken into account to get the correct next-to-eikonal order description of the gluon spectrum. On the other hand, the contribution of $\cR_J$ is structurally very different to the other DBs and can be treated separately . 

As it is thoroughly discussed in \cite{Sadofyev:2021ohn,Kuzmin:2023hko}, after taking medium averages and performing the $q_{1z}$-integral picking up the propagator poles, exponentials lose their dependence on $q_{2z}$ (except for a part of $\cR_G$). Hence, while performing $q_{2z}$-integral, the contour can be closed in either direction and the residues of the poles coming from the scattering potentials must be taken into account explicitly for convergence.

Let us start by the simplest DB diagram $\cR_D$ -- two initial-state scatterings with the medium. The corresponding term in the squared amplitude reads
\begin{align}
\label{Rd initial}
    \langle \cR_D \cR^*_0 \rangle +\text{c.c.} &= i \frac{C^2_F}{N_c} \, g^2  \int_0^L dz \, \rho(z)
    \int_{\x,\q_1,\q_2} \,    \, \frac{4}{\k^2} \, J^*(E,\p - \q_1 - \q_2) J(E,\p) \, \notag
    \\ & \hspace{1.5cm} \times \left[e^{-i(\q_1+\q_2)\cdot \x} \, e^{-i\cQ^-_{p-q_1-q_2}\, z} \, \mathcal{I}_D - e^{i(\q_1+\q_2) \cdot \x} \, e^{i\cQ^-_{p-q_1-q_2} \, z} 
    \, \mathcal{I}_D^* \right] \, ,
\end{align}
where the integral $\mathcal{I}_D$ is defined as 
\begin{align} \label{I_D integral def}
    \mathcal{I}_D = \int \frac{dq_{2z}}{2\pi}\, \frac{E \, v(q_2) \,v(\tilde{q_1})} {(q_{2z}-\cQ^+_{p-q_2}-i\epsilon)(q_{2z}-\cQ^-_{p-q_2}+i\epsilon)}  
\end{align}
with $ \tilde{q}_{1\mu}=(\u\cdot\q_1, \,\q_1, \,-q_{2z}+\cQ^-_{p-q_1-q_2})$. The poles appearing in these two equations can be obtained from \eqref{Qpq poles} by momentum replacements.

Since the nuclear matter is assumed to be homogeneous in the transverse coordinates to the jet, the Fourier factors are the only functions of $\x$. This way, the integration over the transverse coordinates results in $(2\pi)^2\d^{(2)}(\q_1+\q_2)$, which allows removing the integral over one transverse momentum. After that, the integral in \eqref{I_D integral def} can be taken explicitly, resulting in 
\begin{align} \label{I_d integral res}
    \bar{\mathcal{I}}_D \simeq \, i \frac{v(\q)^2}{4} -\frac{\k^2\left((\q\cdot\u)^4+6(\q\cdot\u)^2 R^2 -3 R^4\right)}{64 R^3 \text{x}E}\frac{\partial v^2}{\partial \q^2} \, ,
\end{align}
where $R\equiv\sqrt{\q^2+\mu^2 - (\q\cdot\u)^2}$ comes from the poles of $v(q)$. The bar in \eqref{I_d integral res} indicates that all the terms that vanish under angular averaging of the ingeral $\q$ have been omitted. Thus, the contribution to the squared amplitude reads 
\begin{align}
    &\langle \cR_D \cR^*_0 \rangle + \text{c.c.}  =- \frac{C_F^2}{N_c} \, g^2 \, \frac{2}{\k^2} \int_0^L dz \, \rho(z) \int_{\q} \bigg[\cos\left(\frac{\k^2}{2\text{x}E} \, z\right)   \, \notag
    \\ & \hspace{0.5cm}   - \frac{\k^2\left((\q\cdot\u)^4+6(\q\cdot\u)^2 R^2 -3 R^4\right)}{16 R^3\text{x}E} \frac{1}{v^2}\frac{\partial v^2}{\partial \q^2}  \,  \sin\left(\frac{\k^2}{2\text{x}E} \, z\right)  \bigg] \, v(\q)^2|J(E,\p)|^2  \, .
\end{align}

In the presence of the flow and keeping subeikonal corrections, the real part of the integral \eqref{I_d integral res} is non-zero. This real part couples to the imaginary part of the exponentials of the LPM phases, giving a novel contribution proportional to their sine. A similar behaviour has been observed in \cite{Sadofyev:2021ohn,Kuzmin:2023hko}, where the gradients acting on the LPM phases are the ones coupling to the imaginary part of their exponentials. This will be a common feature for all the DB contributions to the radiation process, as we will see below. 

The next DB term in  the squared amplitude corresponds to the two medium interactions with the final-state quark, and can be written as 
\begin{align}
    \langle \cR_E \cR^*_0 \rangle + \text{c.c.} = -i \frac{C_F^2}{2N_c} \,  g^2 \,  \frac{8}{\k^2} \int_0^L dz\, \rho(z) \int_{\q} [\mathcal{I}_E-\mathcal{I}^*_E] \, |J(E,\p)|^2 \, ,
\end{align}
where we have integrated over the transverse position and one of the transverse momenta to fix $\q\equiv\q_1=-\q_2$, and the integral $\cI_E$ is defined as 
\begin{align}
    \cI_E = \int\frac{dq_{2z}}{2\pi} \, \frac{E \, v(q_2)}{(q_{2z}-\cQ^-_{p-k+q}+i\e)(q_{2z}-\cQ^+_{p-k+q} - i\e)} \bigg[ e^{-i\cQ^-_p z} v\left(q^{ \,p}_{1}\right)-e^{-i\cQ^-_{p-k} z} v\left(q^{\, p-k}_{1}\right)\bigg] \, ,
\end{align}
with $q^{\,p}_{1 \, \mu} =(\q\cdot\u,\q,-q_{2z} + \cQ^-_p)$ and $q^{\, p-k}_{1\,\mu} = (\q\cdot\u,\q,-q_{2z} + \cQ^-_{p-k})$. This integral can be performed explicitly, taking into account both the poles coming from the propagator and the poles coming from the scattering potentials, and reads
\begin{align}
    \bar{\mathcal{I}}_E &\simeq  - i \left(1 - e^{-iQ^-_p z}\right) \frac{v(\q)^2}{4} + e^{iQ^-_p z} \, \frac{\k^2((\q\cdot\u)^4+6(\q\cdot\u)R^2-3R^4)}{64 R^3 \text{x}E} \, \frac{\partial v^2}{\partial \q^2}  \, ,
\end{align}
where all the terms vanishing under angular averaging have been omitted. This way, the $\cR_E$ contribution to the squared amplitude is given by
\begin{align}
    & \langle \cR_E \cR_0^* \rangle + \text{c.c.} = - \frac{C_F^2}{N_c} \, g^2 \, \frac{2}{\k^2} \int_0^L dz\, \rho(z) \int_{\q}  \, \bigg[1-\cos\left(\frac{\k^2}{2\text{x}E} \, z\right) \, \notag
    \\ & \hspace{0.5cm}  - \frac{\k^2((\q\cdot\u)^4+6(\q\cdot\u)^2R^2-3R^4)}{16R^3 \text{x}E} \frac{1}{v^2}\frac{\partial v^2}{\partial \q^2} \, \sin\left(\frac{\k^2}{2\text{x}E} \, z\right)   \bigg] \, v(\q)^2 \,  |J(E,\p)|^2 \, .
\end{align}

The next diagram to be considered is the one with the two interactions in the final-state gluon, where a product of two 3-gluon vertex and two gluon propagators appears. These objects are multiplied by $u_\m$ and $\e^*(k)_{\m}$ which are transverse to $n_\m$, just as the numerator of the gluon propagator $N^{\m\n}$ itself. Thus, the whole object becomes independent of the $z$-components of the momenta and new poles cannot arise from it, exactly as it happens with $\cR_C$. After averaging over the quantum numbers and background field configurations, and performing both $\x$ and $\q_2$ integrals, fixing $\q\equiv\q_1=-\q_2$, $\cR_F$ contribution can be written as 
\begin{align}
     & \langle \cR_F \cR_0^* \rangle + \text{c.c.} = - i \frac{N_c C_F}{2 N_c} \, g^2 \, \frac{8}{\k^2} \, \int_0^L dz \, \rho(z) \int_{\q} \left(1-2\frac{\k\cdot\u}{\text{x}E}\right) \, [\cI_F - \cI^*_F] \, |J(E,\p)|^2 \, ,
\end{align}
where the integral $\cI_F$ is defined as 
\begin{align} \label{I_F integral def}
    \cI_F = \int \frac{dq_{2z}}{2\pi} \frac{\text{x}E \, v(q_2)}{(q_{2z}-\cQ^-_{k+q}+i\e)(q_{2z}-\cQ^-_{k+q}-i\e)} \left[e^{-i\cQ^-_p z} \, v\left(q^{\,p}_{1}\right) - e^{-i\cQ^-_k z} \, v\left(q^{\,k}_{1}\right)\right] \, ,
\end{align}
with  $q^{\,k}_{1\,\mu} = (\q\cdot\u,\q,-q_{2z} + \cQ^-_{k})$. Taking into account the poles coming from the scattering potential, the explicit result of \eqref{I_F integral def} at our working accuracy is 
\begin{align}
    \bar{\mathcal{I}}_F & \simeq  \frac{v(\q)^2}{4}\Bigg[(-i)\left(1 - e^{-iQ^+_p z} \right) \left(1-\frac{\k\cdot\u \, \q^2}{xE} \frac{1}{v^2}\frac{\partial v^2}{\partial \q^2}\right)  \notag
    \\ &  + \frac{1}{16 R^3 xE} \frac{1}{v^2} \frac{\partial v^2}{\partial \q^2} \bigg\{\left(1-e^{-iQ^+_p z}\right) 2\bigg[ (\q\cdot\u)^6 + 5 (\q\cdot\u)^4 R^2 + 3 (\q\cdot\u)^2 R^4-R^6  \notag
    \\ & - \q^2 \,  ((\q\cdot\u)^4+6(\q\cdot\u)^2 R^2 - 3 R^4)\bigg] + e^{-iQ^+_p z} \, \k^2 \left((\q\cdot\u)^4+6(\q\cdot\u)^2 R^2 - 3R^4\right) \bigg\} \Bigg] \,,
\end{align}
where all the terms vanishing under angular averaging have been omitted. After some algebra, $\cR_F$ contribution to the final squared amplitude can be written as 
\begin{align}
     & \langle \cR_F \cR^*_0 \rangle + \text{c.c.} = - C_ F \, g^2 \, \frac{2}{\k^2} \int_0^L dz \, \rho(z) \int_{\q} \left(1-2\frac{\k\cdot\u}{\text{x}E}\right) \left(1-\frac{ \k\cdot\u\, \q^2}{\text{x}E}\frac{1}{v^2}\frac{\partial v^2}{\partial \q^2}\right)  \, \notag
    \\ &\times \Bigg[   1 - \cos\left(\frac{\k^2}{2\text{x}E} \, z\right)   \, \notag
    \\ & \hspace{1cm }+ \frac{(2\mu-\k^2) \,  ((\q\cdot\u)^4+6(\q\cdot\u)^2 R^2 - 3 R^4) -4 R^2 ((\q\cdot\u)^4-R^4)}{16 R^3 \text{x}E} \frac{1}{v^2}\frac{\partial v^2}{\partial \q^2} \, \notag
    \\ & \hspace{3cm} \times \sin\left(\frac{\k^2}{2xE} \, z\right) \Bigg] \, v(\q)^2 |J(E,\p)|^2  \, . 
\end{align}

The following diagram comes from the scattering of the two different final-state particles with the medium. Averaging over quantum numbers and medium configurations, and performing both $\x$ and $\q_2$ integrals in the same way as it has been done in the previous cases, the corresponding term reads
\begin{align}
    \langle \cR_G \cR^*_0\rangle + \text{c.c.} =& ig^2\, \frac{N_c C_F}{4N_c} \frac{8}{\k^2} \int_0^Ldz\, \rho(z) \int_\q   \left(\k\cdot(\k+\q) - \frac{\k^2 \, \u\cdot(\k+\q) + \k\cdot\u \, \q\cdot(\k+\q)}{\text{x}E}\right) \, \notag
    \\ &  \hspace{2cm} \times [\cI_G - \cI_G^*] \, |J(E,\p)|^2 \, ,
\end{align}
where the integral $\cI_G$ is defined as 
\begin{align}
    \mathcal{I}_G &\equiv \int \frac{dq_{2z}}{2\pi} \,\frac{E \, v(q_2)}{(q_{2z}-\cQ^+_{k+q}-i\epsilon)(q_{2z}-\cQ^-_{k+q}+i\epsilon)(q_{2z}-\cQ^-_{p}+\cQ^-_{p-k-q}-i\epsilon)} \, \notag
    \\ & \hspace{1.5cm} \times \Bigg[ \frac{v\left(q^{\,p-k-q}_{1}\right) \, e^{-i(\cQ^-_{p-k-q}+q_{2z}) \, z}}{ q_{2z} + \cQ^-_{p-k-q} - \cQ^+_{p}-i\epsilon} - \frac{ v\left(q^{\, p}_{1}\right) \, e^{-i \cQ^-_{p} \, z}}{-q_{2z} + \cQ^-_{p} - \cQ^+_{p-k-q}-i\epsilon} \Bigg] \, ,
\end{align}
with $q^{\, p-k-q}_{1\,\mu} = (\q\cdot\u,\q,-q_{2z} + \cQ^-_{p-k-q})$. In contrast to the integrals for the previous DB contributions, the Fourier factors in $\cI_G$ have not lost the dependence on $q_{2z}$ completely. Due to the remaining exponential factor, we evaluate the integral by closing the contour below the real axis, obtaining
\begin{align}
    \mathcal{I}_G &\simeq i \left(e^{-i\cQ^-_p z} - e^{-i(\cQ^-_{k+q}+\cQ^-_{p-k-q})z} \right) \frac{v(\q)^2}{2(\k+\q)^2} \left(1-\frac{\q\cdot\u \, \q\cdot(2\k+\q)}{2\text{x}E} \frac{1}{v^2}\frac{\partial v^2}{\partial\q^2}\right) \, \notag
    \\ & + e^{-i\cQ^-_p z} \frac{(\q\cdot\u)^4+6(\q\cdot\u)^2R^2-3R^4}{32 R^3 \text{x}E} \frac{\partial v^2}{\partial\q^2} \, .
\end{align}
where, in order to capture all the $\cO\left(\frac{1}{\tx E}\right)$ corrections of $\cI_G$, the poles must be kept to $\cO\left(\frac{1}{\tx^2 E^2}\right)$. With that result, the contribution of this diagram to the final squared amplitude is given by 
\begin{align}
    \langle R_G R^*_0 \rangle + \text{c.c.} & = C_F  \, g^2\, \int_0^L dz\,  \rho(z) \int_{\q}  \frac{2}{\k^2(\k+\q)^2} v(\q)^2 |J(E,\p)|^2   \, \notag
    \\ & \times \left(\k\cdot(\k+\q) - \frac{\k^2 \, \u\cdot(\k+\q) + \k\cdot\u \, \q\cdot(\k+\q)}{\text{x}E}\right) \, \notag
    \\ & \times \bigg[\left(\cos\left(\frac{(\k+\q)^2-\k^2}{2\text{x}E}\, z\right) -\cos\left(\frac{\k^2}{2\text{x}E}\, z\right)\right)  \left(1-\frac{\q\cdot\u \, \q\cdot(2\k+\q)}{2\text{x}E} \frac{1}{v^2}\frac{\partial v^2}{\partial\q^2}\right) \, \notag
    \\ & \hspace{1cm} - \sin\left(\frac{\k^2}{2\text{x}E}\, z\right) \frac{(\k+\q)^2((\q\cdot\u)^4+6(\q\cdot\u)^2R^2-3R^4)}{16 R^3 \text{x}E} \frac{1}{v^2} \frac{\partial v^2}{\partial\q^2}\bigg] \, .
\end{align}

It is well known that the two diagrams with the first interaction in the initial-state and the second one on a final-state particle do not contribute to the leading eikonal order, and are usually not considered. Moreover, in the scalar `gluon' radiation in $\l \phi^3$ theory, these two diagrams do not contribute to the first subeikonal correction (assuming the LPM phases are small), as it has been shown in \cite{Sadofyev:2021ohn}.  However, both $\cR_H$ and $\cR_I$ will contribute to the first subeikonal correction to the emission of spin-1 gluon, and must be taken into account here.  The term corresponding to the first diagram, where the final-state interacting particle is the quark, reads
\begin{align}
    \langle \cR_H \cR^*_0 \rangle + \text{c.c.} = i g^2 \, \frac{C_F}{4 N_c^2} \, 8 \int_0^L dz \, \rho(z) \, \int_{\q} \left[e^{-i\cQ^-_p z} \cI_H - e^{i\cQ^-_p z} \cI_H^* \right] \, |J(E,\p)|^2 \, .
\end{align}
The latter result has been averaged over quantum numbers and medium configurations, both $\x$ and $\q_2$ integrals have been performed, and the DB integral $\cI_H$ has been defined as 
\begin{align} \label{I_H integral def}
    \cI_H = \int \frac{dq_{2z}}{2\pi} \frac{E}{\text{x}} \frac{v\left(q^{\,p}_{1}\right) v(q_2)}{(q_{2z}-\cQ^-_{p+q}+i\e)(q_{2z}-\cQ^+_{p+q}-i\e)(q_{2z}-\cQ^-_{p-k+q}+i\e)(q_{2z}-\cQ^+_{p-k+q}-i\e)} \, .
\end{align}
Performing the integral in \eqref{I_H integral def} explicitly while taking into account the poles from the scattering potentials, $\cI_H$ can be expressed at our accuracy as 
\begin{align}
    \bar{\cI}_H = - \frac{(\q\cdot\u)^4+6(\q\cdot\u)^2 R^2 -3 R^4}{32 R^3 \text{x}E} \frac{\partial v^2}{\partial\q^2}\, ,
\end{align}
where all the terms vanishing under angular averaging have been omitted. Using this result, the contribution of $\cR_H$ to the final squared amplitude is 
\begin{align}
     \langle \cR_H \cR^*_0\rangle + \text{c.c.} &= \frac{C_F}{N^2_c} \, g^2 \int_0^L dz \, \rho(z) \, \int_{\q} \frac{(\q\cdot\u)^4+6(\q\cdot\u)^2 R^2 -3 R^4}{8 R^3 \text{x} E}  \frac{\partial v^2}{\partial \q^2} \, \sin\left(\frac{\k^2}{2xE} \, z\right) |J(E,\p)|^2  \, .
\end{align}

The other contribution of this class is given by $\cR_I$, where the final-state interacting particle is the gluon. As we discussed computing $\cR_C$ and $\cR_F$, the product of the gluon propagator with the 3-gluon vertex, the external field and the final polarization vector cannot introduce any new pole. Thus, after averaging over quantum numbers and medium configurations, and performing both $\x$ and $\q_2$ integrals, the corresponding term reads 
\begin{align}
    \langle \cR_I \cR_0^* \rangle + \text{c.c.} =&  -i g^2 \, \frac{C_F N_c}{4N_c} \frac{8}{\k^2} \int_0^L dz \, \rho(z) \int_\q \k\cdot(\k+\q) \, \left[e^{-i\cQ^-_p z} \cI_I - e^{i\cQ^-_p z} \cI_I^*\right] \, |J(E,\p)|^2 \, , 
\end{align}
where the integral $\cI_I$ is defined as 
\begin{align} \label{I_I integral def}
    \cI_I = \int\frac{dq_{2z}}{2\pi} \frac{E \, v\left(q^{\, p}_{1}\right) v(q_2)}{(q_{2z}-\cQ^-_{p+q}+i\e)(q_{2z}-\cQ^+_{p+q}-i\e)(q_{2z}-\cQ^-_{k+q}+i\e)(q_{2z}-\cQ^+_{k+q}-i\e)} \, .
\end{align}
Taking into account the poles coming from the scattering potential, the explicit result of \eqref{I_I integral def} at our working accuracy is
\begin{align}
    \cI_I \simeq i \frac{\q\cdot\u}{4\text{x}E} \frac{\partial v^2}{\partial \q^2} - \frac{(\q\cdot\u)^4+6(\q\cdot\u)^2 R^2 -3 R^4}{32 R^3 \text{x}E} \frac{\partial v^2}{\partial\q^2}\, .
\end{align}
With that, $\cR_I$ contribution to the squared amplitude can be written as
\begin{align}
     &\langle \cR_I \cR^*_0\rangle + \text{c.c.} = C_F \,  g^2  \int_0^L dz \, \rho(z) \int_{\q}    \, \notag
     \\ & \times \left[\frac{\k\cdot\u \, \q^2}{2\k^2} \, \cos\left(\frac{\k^2}{2xE} \, z\right) - \frac{(\q\cdot\u)^4+6(\q\cdot\u)^2 R^2 -3 R^4}{8 R^3} \sin\left(\frac{\k^2}{2xE} \, z\right) \right] \frac{1}{\tx E} \frac{\partial v^2}{\partial \q^2}   |J(E,\p)|^2  \, .
\end{align}

The final DB contribution to the squared amplitude comes from $\cR_J$, where the final-state gluon interacts twice with the medium via a 4-gluon vertex, instead of two 3-gluon vertices as in $\cR_F$. Similarly to the previous two diagrams, $\cR_J$ is usually not considered in the literature because it is subdominant in energy. However, it contributes at the first subeikonal order and must be studied to get a complete description of the soft gluon emission at N=1. 

The amplitude of this diagram is given by 
\begin{align}
    i \cR_J =& \int \frac{d^4q_1}{(2\pi)^4} \frac{d^4q_2}{(2\pi)^4}  \, [igt^a (2p-k-q_1-q_2)_{\m'}] \,  \frac{(-i)N^{\m'\m}(k-q_1-q_2)}{(k-q_1-q_2)^2+i\e} \, V^{abcd}_{\m\n\rho\s} \, \notag
    \\ & \times \e^{*\s}(k) \, A^{c\rho}(q_2) \, A^{b\n}(q_1) \,   \frac{i}{(p-q_1-q_2)^2+i\e} J(p-q_1-q_2) \, ,
\end{align}
where the 4-gluon vertex is defined as 
\begin{align}
    V^{abcd}_{\m\n\rho\s} = -i g^2 [&f^{abe}f^{cde}(g^{\m\rho}g^{\n\s}-g^{\m\s}g^{\n\rho}) + f^{ace}f^{bde}(g^{\m\n}g^{\rho\s}-g^{\m\s}g^{\n\rho}) \, \notag
    \\ & + f^{ade}f^{bce}(g^{\m\n}g^{\rho\s}-g^{\m\rho}g^{\n\s})] \, .
\end{align}

Similarly to what is done for the other DB diagrams, we must multiply by the vacuum amplitude $\cR_0$ and average  over all possible configurations of the background field. This average will bring a factor $\d^{bc}$, which forces two colors of the 4-gluon vertex to be the same, simplifying its structure. Since the numerator of the gluon propagator and the background fields are orthogonal to the axial vector $n$, the tensor structure is independent of the z-component of the momenta and cannot bring any new poles. Averaging over initial and summing over final quantum numbers, and performing both $\x$ and $\q_2$ integrals, the corresponding term reads
\begin{align}
    \langle \cR_J \cR^*_0 \rangle + \text{c.c.} =& - ig^2 \, 2 C_F\,  \, \frac{\k^2 \, (1-\u^2) + (\k\cdot\u)^2}{\k^4 \text{x} E} \int_0^L dz \, \rho(z) \int_\q \, \notag
    \\ & \times\left[\left(e^{-i\cQ^-_p z} - 1\right)\cI_J - \left(e^{i\cQ^-_p z} - 1\right)\cI^*_J\right] \, |J(E,\p)|^2 \, , 
\end{align}
where the integral is defined as 
\begin{align}
    \cI_J = \int \frac{dq_{2z}}{2\pi} \, v(\tilde{q}_1) \,  v(q_2) = \frac{(R^2+(\q\cdot\u)^2)^2}{4 R^3} \, v(\q)^2 \, ,
\end{align}
with $\tilde{q}_{1\m}= (\q\cdot\u,\q,-q_{2z})$. Thus, the contribution of the 4-gluon vertex diagram to the final squared amplitude is 
\begin{align}
    \langle \cR_J \cR^*_0\rangle + \text{c.c.} =& C_F \,  g^2 \, \frac{\k^2 \, (1-\u^2) + (\k\cdot\u)^2}{\k^4 \, \tx E} \int_0^L dz \, \rho(z) \int_\q \, \notag
    \\ & \times \frac{\left(R^2+(\q\cdot\u)^2\right)^2}{R^3} \, \, v(\q)^2 \, \sin\left(\frac{\k^2}{2\text{x}E}\, z\right) \, |J(E,\p)|^2 \, .
\end{align}

Working with scalar quarks, there is a seagull vertex in the theory that may be needed to get a correct description of the process. Diagrams containing this vertex are usually omitted in scalar QCD jet quenching calculations because they do not contribute to the leading eikonal order, see e.g. the discussion in \cite{Sadofyev:2021ohn}. An explicit calculation shows that their amplitudes are of the order $\cO\left(\frac{1}{E} \left(\frac{1}{\tx E}z\right)^n\right)$, which is beyond the accuracy of our consideration. Hence, there are no contributions coming from these seagull diagrams to our result in this paper.

\subsection{Final parton distribution and its properties}

The final state parton distribution can be expressed through the squared amplitude of the process as
\begin{align}
     E\, \frac{d\cN^{(1)}}{d^2k \, d\tx \, d^2p \, dE} \equiv \frac{1}{\left[2(2\pi)^3\right]^2} \, \frac{1}{\tx} \, \langle|\cR_{N=1}|^2\rangle \,,
\end{align}
where  $\cR_{N=1}$ is the full amplitude at first order in the opacity expansion, and we have included the corresponding final-state phase factor in the soft gluon limit. 

The initial parton distribution is defined through the source as $E\, \frac{d\cN^{(0)}}{d^2p \, dE}\equiv \frac{1}{2(2\pi)^3} |J(E,\p)|^2$. Hence, the broad source approximation, which assumes $J$ to be a slowly varying function of the transverse momenta, implies that $E\, \frac{d\cN^{(0)}}{d^2(p-q)\, dE} \simeq E\, \frac{d\cN^{(0)}}{d^2p \, dE}$. This way, the final parton distribution reads
\begin{align}\label{dN^(1) final}
    &E\, \frac{d\cN^{(1)}}{d^2k \, d\tx \, d^2p \, dE} = \frac{1}{2(2\pi)^3} \, \frac{1}{\tx}\left(E\, \frac{d\cN^{(0)}}{dE \, d^2p}\right) \, C_F \, g^2 \int_0^L dz\, \rho(z) \int_\q v(\q)^2 \, \notag
    \\& \times \Bigg\{ \Bigg[4 \, \frac{\k\cdot\q}{\k^2(\k-\q)^2}-\frac{2}{\tx E} \frac{1}{\k^2(\k-\q)^2} \bigg(2(\k-\q)\cdot\u \, \k^2 +2 \, \k\cdot\u \, (\k-\q)\cdot\q \, \notag
    \\ & \hspace{1.5cm} + \k\cdot\q \, \q\cdot\u \, (2\k^2-(\k-\q)^2) \frac{1}{v^2}\frac{\partial v^2}{\partial \q^2}\bigg) \Bigg] \left(1-\cos\left(\frac{(\k-\q)^2}{2\tx E}\,z\right)\right) \, \notag
    \\ &  + \frac{1}{\tx E} \frac{\k\cdot\u}{\k^2} \left(4+\q^2 \, \frac{1}{v^2}\frac{\partial v^2}{\partial \q^2}\right) \left(1-\cos\left(\frac{\k^2}{2\tx E}\,z\right)\right) \, \notag
    \\ & - \frac{1}{4 R^3 \tx E} \frac{\mu^2((\q\cdot\u)^4 + 6(\q\cdot\u)^2 R^2-3R^4) - 2R^2((\q\cdot\u)^4-R^4)}{\k^2} \, \frac{1}{v^2}\frac{\partial v^2}{\partial \q^2} \,  \sin\left(\frac{\k^2}{2\tx E}\,z\right) \, \notag
    \\ & +\frac{1}{\tx E}\frac{\k^2 \, (1-\u^2) + (\k\cdot\u)^2}{\k^4} \, \frac{ \left(R^2+(\q\cdot\u)^2\right)^2}{R^3} \, \sin\left(\frac{\k^2}{2\tx E}\,z\right) \Bigg\} \, ,
\end{align}
where the LO term can be easily identified as the usual GLV result under the soft gluon approximation, see e.g. \cite{Gyulassy:2000er}.

The final state parton distribution \eqref{dN^(1) final} obtained in this paper extends the GLV result for the radiation of a spin-1 gluon at N=1 under the soft approximation by including the novel subeikonal corrections in flowing nuclear matter. Going to the limit considered in \cite{Sadofyev:2021ohn}, one can check that the spin-1 spectrum is indeed related with the scalar emission spectrum under the light-front wave function replacement up to the first subeikonal order. However, one should keep all the diagrams at the given accuracy before averaging them over the momentum angles since the light-front wave functions of different theories may behave differently.

Given the explicit form \eqref{dN^(1) final}, it is obvious that the final state parton distribution can be factorized into the initial quark distribution and the medium-induced emmission spectrum, which is defined as 
\begin{align}
    \tx E \, \frac{d\cN^{(1)}}{d^2k \, d\tx \, d^2p \, dE } \equiv \tx\, \frac{d\cI^{(1)}}{d\tx d^2k} \, E \, \frac{d\cN^{(0)}}{dE d^2p} \, .
\end{align}

In order to learn more about the medium-induced spectrum, let us first start with the static matter limit $\u=0$. Then, the only non-vanishing subeikonal corrections are the ones proportional to the sine of the LPM phase coming from the DB diagrams. Hence, the gluon spectrum reads
\begin{align} \label{dI u=0 z integ}
     \tx \, \frac{d\cI^{(1)}}{d\tx d^2k} \bigg|_{\u=0} =& \frac{C_F \, g^2}{2(2\pi)^3}\int_0^L dz\, \rho(z) \int_\q v(\q)^2 \Bigg\{ 4 \, \frac{\k\cdot\q}{\k^2(\k-\q)^2} \left(1-\cos\left(\frac{(\k-\q)^2}{2\tx E}\,z\right)\right) \, \notag
     \\ & + \frac{1}{2\tx E} \frac{1}{\k^2} \left(\frac{3}{2}\mu^2 R_0 - 2 R^3_0\right) \, \frac{1}{v^2}\frac{\partial v^2}{\partial \q^2} \,\sin\left(\frac{\k^2}{2\tx E}\,z\right) \Bigg\} \, , 
\end{align}
with $R_0\equiv R|_{\u=0}=\sqrt{\q^2 + \mu^2}$. One should notice that the form of the corrections in \eqref{dI u=0 z integ} is governed by the details of the particular potential $v(q)$. Following \cite{Gyulassy:2000fs,Gyulassy:2000er,Kuzmin:2023hko}, we can treat the $z$ integral analytically by choosing a smooth longitudinal profile for the source density $\rho(z)=2 \rho_0 \, e^{-\frac{2z}{L}}$. Then, the spectrum is given by
\begin{align} \label{dI u=0 plot}
    \tx \, \frac{d\cI^{(1)}}{d\tx d^2k} \bigg|_{\u=0} =& \frac{4 \, \a_s \, \chi \, N_c \, \mu^2 \, L}{\pi} \int_\q \frac{1}{(\q^2+\mu^2)^2}\Bigg\{ 2 \, \frac{\k\cdot\q}{\k^2} \frac{L \, (\k-\q)^2 }{L^2(\k-\q)^4+16 \tx^2 E^2} \, \notag
     \\ & - \left(3\mu^2 R_0 - 4 R^3_0\right) \, \frac{1}{\q^2+\mu^2} \, \frac{1}{L^2 \k^4 + 16 \tx^2 E^2} \Bigg\} \, ,
\end{align}
where the opacity $\chi = \frac{C_F g^4 \rho_0}{2 N_c 4 \pi \mu^2}L$ has been introduced. Notice that the leading subeikonal correction in the expression above appears of the same order as the eikonal term. However, it is not surprising since the eikonal term gains additional length enhancement and still dominates the spectrum.

\begin{figure}[t!]
    \centering
    \includegraphics[width=1\textwidth]{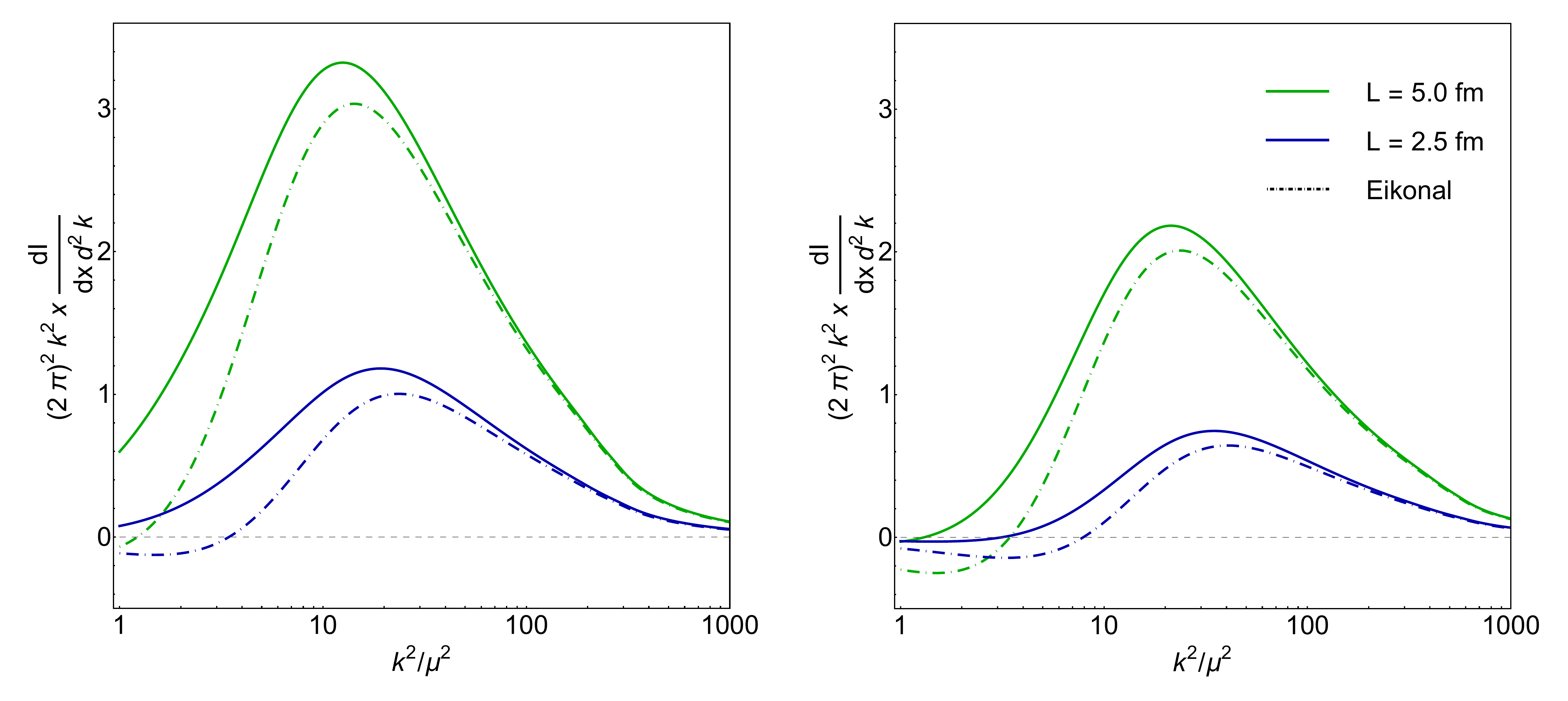}
    \vspace*{-10mm}\caption{The medium-induced spectrum is plotted for two energies $E=50 \, \text{GeV}$ (left) and $E=100 \, \text{GeV}$ (right), and two medium lengths $L=5 \, \text{fm}$ (green) and $L=2.5 \, \text{fm}$ (blue). The dashed lines correspond to the spectrum at leading eikonal order (GLV), while the solid lines include the subeikonal corrections to the medium-induced spectrum for static matter. }
\label{f:LO + corr}
\end{figure}

In Fig.~\ref{f:LO + corr}, we plot the spectrum \eqref{dI u=0 plot} for two energies $E = 50 \, \text{GeV}$ (left) and $E = 100 \, \text{GeV}$ (right), keeping the mean free path $\lambda = \frac{L}{\chi}$ fixed and assuming $\chi$=3 at $L=5\, \text{fm}$. Both the leading eikonal part and the full spectrum \eqref{dI u=0 plot}  are shown to emphasise the modifications due to the subeikonal corrections. We set $\a_s=0.3$, $\mu=0.6\, \text{GeV}$ and $\tx=0.1$. Fig.~\ref{f:LO + corr} proves that subeikonal corrections to the spectrum may become relevant for gluons emitted with small transverse momentum $\k$ even at moderate energies.

The same analytical treatment of the $z$ integral can be applied to the case of flowing matter, with the medium-induced spectrum being
\begin{align}\label{dI u plot}
    & \tx \, \frac{d\cI^{(1)}}{d\tx d^2k} = \frac{2 \, \a_s \, \chi \, N_c \, \mu^2 \, L}{\pi} \int_\q \frac{1}{(\q^2+\mu^2)^2}\, \notag
    \\ & \times \Bigg\{ \Bigg[4 \frac{\k\cdot\q}{\k^2}- \frac{2}{\tx E} \frac{1}{\k^2} \bigg(2(\k-\q)\cdot\u \, \k^2 +2 \, \k\cdot\u \, (\k-\q)\cdot\q \, \notag
    \\ & \hspace{1.5cm} + \k\cdot\q \, \q\cdot\u \, (2\k^2-(\k-\q)^2) \frac{-2}{\q^2+\mu^2} \bigg) \Bigg] \frac{L \, (\k-\q)^2}{L^2 (\k-\q)^4+16 \tx^2 E^2}  \, \notag
    \\ &  + \frac{1}{\tx E} \, \k\cdot\u \, \left(4+\q^2 \, \frac{-2}{\q^2+\mu^2}\right)\frac{L \, \k^2}{L^2\k^4+16 \tx^2 E^2} \, \notag
    \\ & - \frac{1}{R^3} \frac{ \mu^2((\q\cdot\u)^4 + 6(\q\cdot\u)^2 R^2-3R^4) - 2R^2((\q\cdot\u)^4-R^4)}{L^2 \k^4 + 16 \tx^2 E^2} \, \frac{-2}{\q^2+\mu^2} \, \notag
    \\ & +\frac{\k^2 \, (1-\u^2) + (\k\cdot\u)^2}{\k^2 \, R^3} \, (\q^2+\mu^2) \, \frac{4}{L^2\k^4+16 \tx^2 E^2} \Bigg\} \, . 
\end{align}
\begin{figure}[h!] 
    \centering
    \begin{subfigure}
        \centering
        \includegraphics[width=1\textwidth]{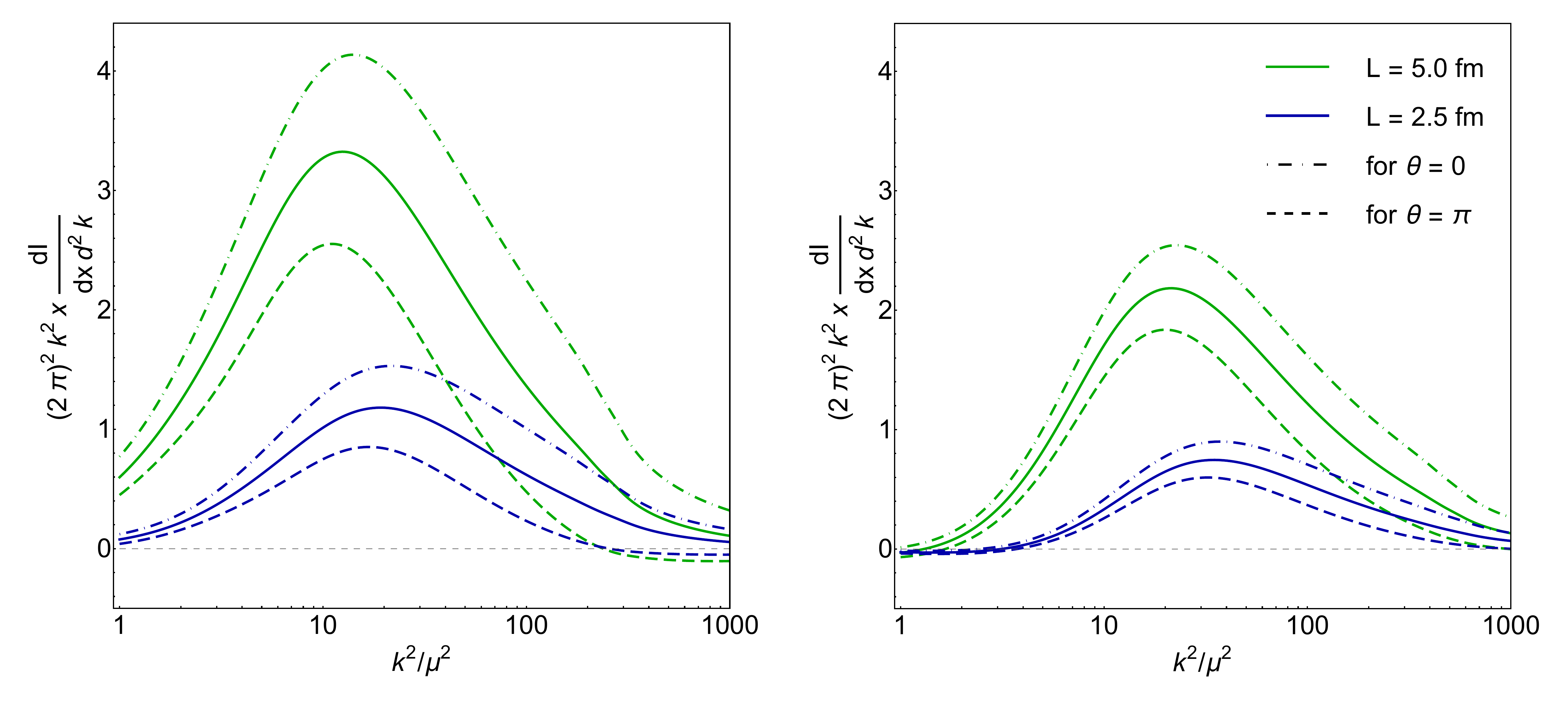}
    \end{subfigure}
    \\
    \begin{subfigure}
        \centering
        \includegraphics[width=1\textwidth]{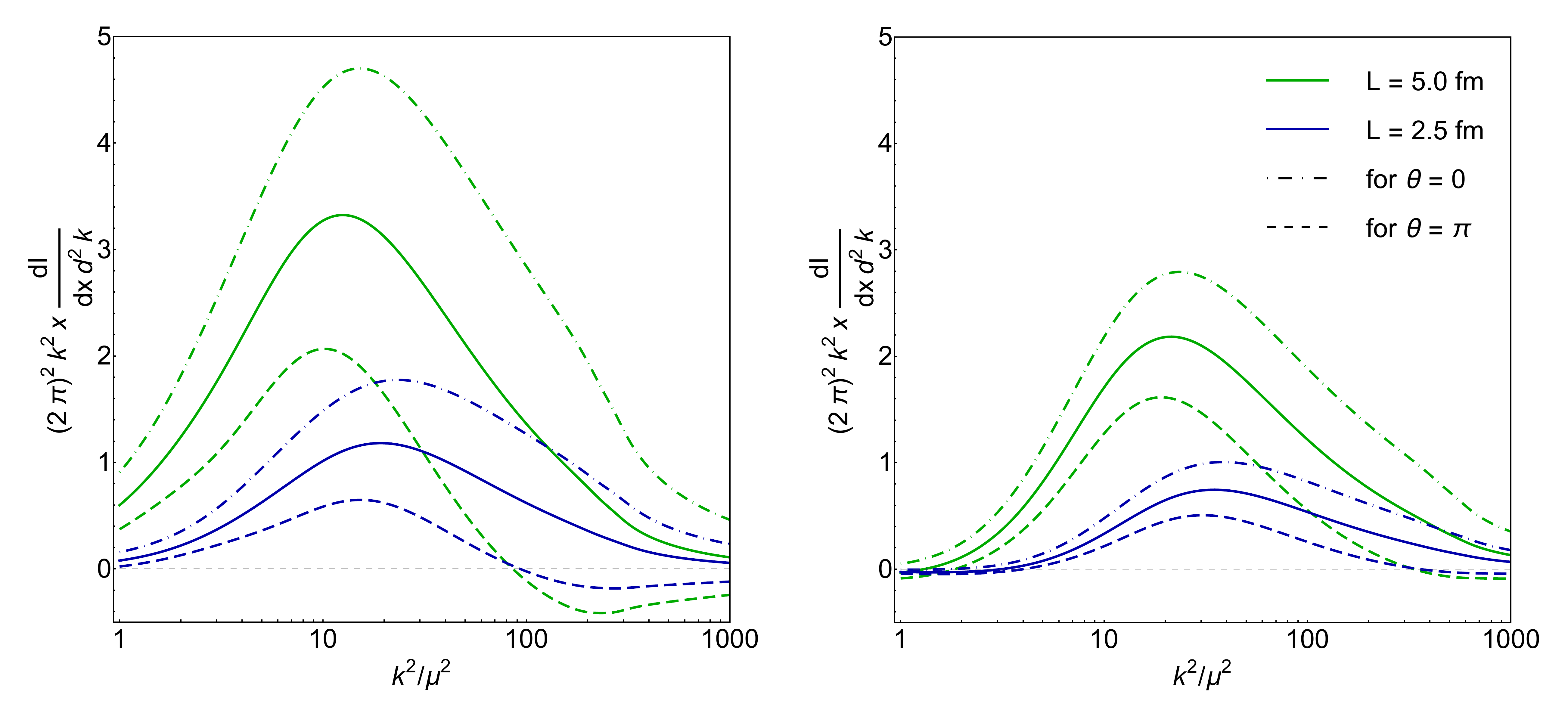}
    \end{subfigure}
    \vspace*{-10mm}\caption{The medium-induced spectrum is plotted for two energies $E = 50 \, \text{GeV}$ (left column) and $E = 100 \, \text{GeV}$ (right column), and two magnitudes of the transverse flow $|\u|=0.3$ (top row) and $|\u|=0.5$ (bottom row). The mean free path $\lambda=\frac{L}{\chi}$ is kept fixed assuming $\chi=3$ at $L=5\, \text{fm}$, and two different lengths $L=5\,\text{fm}$ (green) and  $L=2.5\,\text{fm}$ (blue) of the medium are plotted. The solid line corresponds to the static limit $\u=0$, while the dashed lines correspond to $\k$ and $\u$ being parallel ($\theta = 0$) or antiparallel ($\theta = \pi$).}
\label{f:dI with u eff}
\end{figure}

In Fig.~\ref{f:dI with u eff}, we plot the spectrum \eqref{dI u plot} for two energies $E = 50 \, \text{GeV}$ (left column) and $E = 100 \, \text{GeV}$ (right column), and two magnitudes of the transverse flow $|\u|=0.3$ (top row) and $|\u|=0.5$ (bottom row). The mean free path $\lambda = \frac{L}{\chi}$ is kept fixed assuming $\chi$=3 at $L=5\, \text{fm}$, and the spectrum is plotted for $L=5\,\text{fm}$ and $L=2.5\,\text{fm}$. The remaining parameters are set to the values used in Fig.~\ref{f:LO + corr}. It is easy to see from Fig.~\ref{f:dI with u eff} that the angle $\theta$ between $\u$ and $\k$ regulates the effect of the flow correction to the spectrum. If the transverse momentum of the gluon and the flow are parallel on the transverse plane ($\theta=0$), the gluon radiation gets enhanced, while if they are antiparallel ($\theta=\pi$) it gets depleted. Thus, one can affirm that the final transverse momentum of the emitted gluons tends to align along the flow direction. Notice that the dependence of the spectrum on the transverse flow is not just through $\k\cdot\u$, but also $\u^2$ and $(\q\cdot\u)^2$. Thus, the conditions of static matter and $\k$ and $\u$ being orthogonal vectors, forcing $\k\cdot\u=0$, are not equivalent. Consequently, the curve at $\theta=\frac{\pi}{2}$ would not coincide with the solid line but would correspond to a new one laying between the two dashed lines. The change in shape and magnitude of the spectrum plotted in Fig.~\ref{f:dI with u eff} shows that the flow corrections become significant even at moderate energies and flow velocities, especially for larger systems.

\section{Discussion and conclusions}

In this work, we have studied the effect of the flow on the branching process of a highly energetic quark, obtaining the double differential medium-induced  spectrum of soft gluons in the presence of a flowing homogenous medium. The spectrum is computed at first order in the opacity expansion, including the leading subeikonal corrections for a real spin-1 gluon, extending the formalism developed in \cite{Sadofyev:2021ohn,Kuzmin:2023hko}. Apart from showing the analytical form of the spectrum \eqref{dN^(1) final}, we have evaluated it for different energies, medium lengths and flow velocities.

As it has been seen above, the subeikonal corrections do not vanish in the case of static matter, enhancing the spectrum substantially at low transverse momenta of the emitted gluon. In the presence of transverse flow, both the shape and magnitude of the spectrum get modified. The angle $\theta$ between the final transverse momentum of the gluon $\k$ and the transverse flow $\u$ modulates whether the spectrum gets enhanced or depleted, while it is observed that $\k$ tends to align along the direction of the flow. The dependence of the spectrum on the transverse flow is not linear, and therefore $\theta=\frac{\pi}{2}$ does not coincide with the static case, in contrast to what has been observed for the gradient corrections, see e.g. \cite{Barata:2023qds,Kuzmin:2023hko}. 

The presented results contribute to the ongoing efforts to improve the theory of jet-medium interactions, and, moreover, could be extended in multiple ways. First, it would be interesting to get the subeikonal flow corrections for the medium-induced spectrum in the dense regime, resumming multiple scatterings, although such a calculation will be challenging. In addition, the theoretical framework developed in this paper could be implemented into phenomenological considerations to provide a more realistic description of the jet interaction with the evolving QGP, \textit{c.f.} the discussions in \cite{Barata:2023zqg,Barata:2023bhh,Barata:2023vnl,Antiporda:2021hpk,He:2020iow,Andres:2023xwr}. For instance, having the spectrum with flow in hand, one can study the sensitivity of jet angularities and other shape observables to the flow, starting with the simplest geometries, see e.g. \cite{Barata:2023zqg}.

\section*{Acknowledgments}

The authors are grateful to J. Barata, F. Dominguez and C. Salgado for multiple discussions and comments on this work. The authors would like to particularly thank A.V. Sadofyev for his insightful observations that have been essential to shape this project. The work of XML is supported by European Research Council project ERC-2018-ADG-835105 YoctoLHC; by Xunta de Galicia (Centro singular de investigación de Galicia accreditation 2019-2022), by European Union ERDF; and by Grant CEX2023- 001318-M funded by MICIU/AEI/10.13039/501100011033 and by ERDF/EU. XML contribution to this work is also supported under scholarship No. PRE2021-097748, funded by MCIN/AEI/10.13039/501100011033 and FSE+.

\bibliographystyle{bibstyle}
\bibliography{references}

\end{document}